\documentclass[prl,twocolumn,showpacs,preprintnumbers,amsmath,amssymb]{revtex4}
\usepackage{graphicx}
\begin{document}

%hc \title{Possible structures for Si and Ge sub-nano-wires}
\title{Structures of Si and Ge nanowires in the sub-nanometer range}
\author{R. Kagimura}
\author{R. W. Nunes}
\author{H. Chacham}

\affiliation{Departamento de F\'{\i}sica, ICEX, Universidade Federal de 
Minas Gerais, CP 702, 30123-970,Belo Horizonte, MG, Brazil.}

\date{\today}

\begin{abstract}

We report an {\it ab-initio} investigation of several possible Si and
Ge pristine nanowires with diameters between 0.5 and 1.2 nm.  
We considered nanowires based on the diamond structure,
high-density bulk structures, and fullerene-like structures.
We find that the diamond structure nanowires are unstable
for diameters smaller than 1 nm, and
undergo considerable structural transformations towards
amorphous-like wires. Such instability is consistent with a
continuum model that predicts, for both Si and Ge, a stability crossover 
between diamond and high-density-structure nanowires for diameters smaller 
than 1 nm. For diameters between 0.8 nm and 1 nm, filled-fullerene wires
are the most stable ones.  For even smaller diameters (d 
$\sim$ 0.5 nm), we find that a simple hexagonal structure is particularly
stable for both Si and Ge.

\end{abstract}

% 61.46.+w Structure of solids and liquids; crystallography 
% Nanoscale materials: clusters, nanoparticles, nanotubes, and nanocrystals

\maketitle

Semiconductor nanowires with diameters of a few nanometers can be
grown nowadays by vapor-liquid-solid ~\cite{science98,apl98},
solution-growth~\cite{science00}, or oxide-assisted growth
methods~\cite{science03}. These nanowires usually depict a crystalline
core surrounded by an oxide outer layer.  Further removal of the oxide
layer by acid treatment may lead to hydrogen-passivated 
silicon nanowires as thin as one
nanometer~\cite{science03}. 
Pristine (non-passivated) silicon wires with diameters of a few
nanometers have also been produced from Si vapor deposited on graphite
~\cite{fulerexp}.
The elongated shape of silicon and germanium clusters of up to a few
tens of atoms, determined by mobility measurements ~\cite{prl92,prl94}, 
indicates that even thinner pristine structures, with diameters smaller
than 1 nm, can been produced.

The growth of
such small-diameter structures raises the question of the limit of a
bulk-like description of bonding in these nanowires, since for small
enough diameters the predominance of surface atoms over inner
(bulk-like) atoms will eventually lead to bonds (and structures)
distinct from those of the bulk system.
In the present work, we use first principles calculations to
investigate several periodic structures of silicon and germanium
pristine nanowires of infinite length,  
with diameters ranging from 0.45 to 1.25 nm.  
The nanowire structures considered are based on the diamond structure,
fullerene-like structures, and the high-density bulk structures
$\beta$-tin, simple cubic (sc), and simple hexagonal (sh). 

Our calculations are performed in the framework of the density
functional theory \cite{kohn}, within the generalized-gradient
approximation (GGA)~\cite{gga} for the exchange-correlation energy
functional, and the soft norm-conserving pseudopotentials of
Troullier-Martins \cite{pse-tm} in the Kleinman-Bylander factorized
form \cite{KL}.  We use a method \cite{siesta} in which the one
electron wavefunctions are expressed as linear combinations of
pseudo-atomic numerical orbitals of finite range. A double-zeta basis
set is employed, with polarization orbitals included for all atoms. 
For the nanowire calculations, we employ supercells that are periodic
along the wire axis, and that are wide enough in the perpendicular
directions to avoid interaction between periodic images.
All the geometries were optimized until residual forces were
less than 0.04 eV/\AA.  Total-energy differences were converged to
within 4 meV/atom with respect to orbital range and k-point sampling.

Most of the nanowire structures we consider in this work are derived
from crystalline structures. At zero pressure and temperature, the
structure with the lowest energy per atom is the cubic diamond (cd),
for both Si and Ge, and we use it as the reference structure in our
calculations. As a first test of the methodology we employ, we compute
the total energy per atom ($E_{tot}$), at zero pressure, for the
following bulk phases: cd, hexagonal diamond (hd), $\beta$-tin, sh,
sc, bcc, hcp, and fcc.  In table~I, we show the total energy per atom
of each structure, $\Delta E_{tot}= E_{tot}-E_{tot}^{cd}$, relative to
the total energy of the cd phase, $E_{tot}^{cd}$. We observe that
$\Delta E_{tot}$ is within 0.20-0.45 eV/atom for the sc, sh, and
$\beta$-tin phases, for both Si and Ge. Our $\Delta E_{tot}$ results
for the sc, sh, and $\beta$-tin phases of Si, and for the $\beta$-tin
phase of Ge are in good agreement with recent GGA
calculations~\cite{phase03,phase95}.  As a further test of the
reliability of our calculations, we consider the structural transition
from diamond to $\beta$-tin, that occurs for both Si and Ge under
pressure.  The experimentally measured transition pressure is 117 kBar
for Si and 106 kBar for Ge~\cite{phasereview}. 
By computing the enthalpy $H = E + pV$ from constant pressure
calculations, we obtained theoretical values of 124 kBar and 103 kbar
for Si and Ge, respectively, both in very good agreement with the
experimental values. These results also agree with previous GGA
calculations~\cite{phase95,phase04,lda}.

\begin{table}
\centering
%\caption{Relative total energy per atom ($\Delta E_{tot}$), in eV/atom, 
\caption{Calculated total energies per atom ($\Delta E_{tot}$), in eV/atom, 
of selected Si and Ge bulk phases, 
relative to the cubic diamond phase.} 
\vspace{0.30cm}
\label{phases}
\begin{tabular}{lccccccc} \hline \hline
                      &
%hc \multicolumn{1}{c}{cd}&
%rk\multicolumn{1}{c}{hd}&
%rk\multicolumn{1}{c}{$\beta$-tin}&
%rk\multicolumn{1}{c}{sh}&
%rk\multicolumn{1}{c}{sc}&
%rk\multicolumn{1}{c}{bcc}&
%rk\multicolumn{1}{c}{hcp}&
%rk\multicolumn{1}{c}{fcc}\\
\multicolumn{1}{l}{hd}&
\multicolumn{1}{l}{$\beta$-tin}&
\multicolumn{1}{l}{sh}&
\multicolumn{1}{l}{sc}&
\multicolumn{1}{l}{bcc}&
\multicolumn{1}{l}{hcp}&
\multicolumn{1}{l}{fcc}\\
\hline
%--------------------------------------------------------------------%
%hc Si                     \\
%hc $\Delta E_{tot}$    & 0.000  &0.096   & 0.388  &0.395
Si \:\:  &0.096 \:\:  & 0.388 \:\: &0.395 \:\:
& 0.442 \:\: & 0.610 \:\: &  0.652 \:\: & 0.673 \:\: \\
\hline
%hc Ge                     \\
%hc $\Delta E_{tot}$  & 0.000  &0.021  & 0.240 & 0.236 & 0.236  & 0.296 & 0.325
Ge  \:\: &0.021\:\:  & 0.240 \:\:& 0.236\:\: & 0.236\:\:  & 0.296\:\: 
& 0.325\:\: & 0.315 \:\:  \\

\hline
\hline
%---------------------------------------------------------------------%
\end{tabular}
\end{table}

\begin{figure}[!t]
\includegraphics{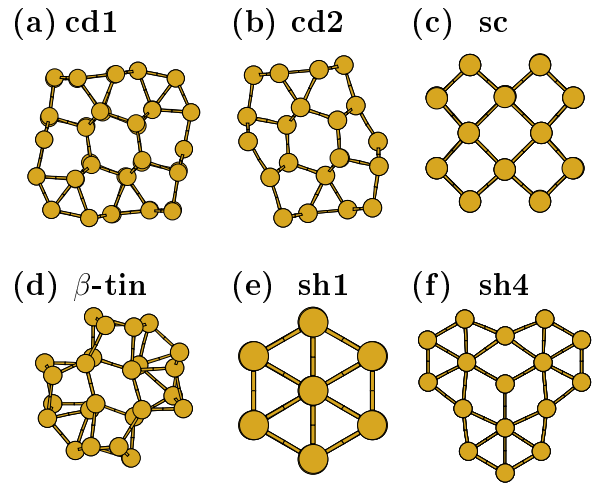}
\caption{Cross sections of selected Ge nanowire structures labeled 
according to the
parent bulk phase. In (a) and (b), wires derived from the
cubic diamond structure, with axis along (110) direction; in (c), a
simple cubic wire; in (d), a $\beta-tin$ wire with axis along bulk
$c$-direction; in (e) and (f), simple hexagonal wires with axis
along bulk $c$-direction. }
\label{nanowires}
\end{figure}

\begin{figure}[!t]
\includegraphics[height=5.5cm]{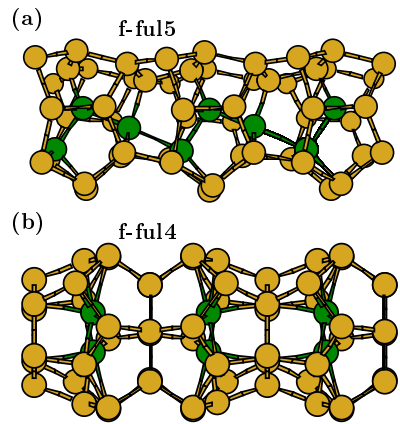}
\caption{Side view of corrugated Ge nanowire structures. In (a), corrugated
wire resulting from a structural instability of a (100) cubic diamond
nanowire; in (b), a filled-fullerene nanowire. Inner atoms are shown as
green spheres.}
\label{nanowires2}
\end{figure}

We now address the structure and energetics of several stable
structures of Si and Ge nanowires with diameters between  0.45 and 1.25 nm. 
Three classes of structures are considered. The first one 
derives from the cubic diamond (cd) bulk phase, with the nanowire axis 
oriented along (100) or (110). The latter corresponds to the usual
orientation of observed Si nanowires with diameters between 3 and 10 nm 
~\cite{nano04}.  
The second class of structures derives from fullerene-like structures
~\cite{parrinello,fulerexp,fulerteo}.
The third class of structures derives from the high-density
$\beta$-tin, sc, and sh bulk phases. 
Our motivation for the study of the high-density structures 
initially came from the observation, in our {\it ab initio} calculations,
of a structural instability of a (6,0) silicon nanotube which
spontaneously deformed into a simple-cubic structure of considerably
low formation energy, shown in 
Fig.~\ref{nanowires}(c). This finding led us to consider the
$\beta$-tin and the sh high-density phases, 
which also have relatively small
formation energies in bulk, as shown in table I. In the following, we
describe the three classes of structures:

{\it Diamond-structure nanowires -} We considered several
nanowires based on the cd
structure with diameters between 0.5 nm and 1.2 nm, 
oriented either along (110) or (100).
These structures were obtained from the cd bulk by defining
the wire axis along the indicated crystalline direction, and by including
atoms that fall within a specified distance from the axis. From this
initial geometry, we removed low-coordinated surface atoms and performed
geometry optimization using the {\it ab initio} scheme. For both Si and
Ge, only the two
widest wires, oriented along (110), remained diamond-like after
geometry optimization. 
The geometries of these wires, labeled as cd1 and cd2, are shown in 
Figs.~\ref{nanowires} (a) and (b) in the case of Ge. The corresponding 
structures for Si are very similar.  Note that both the cd1 and cd2 wires
undergo reconstruction at the surface but retain a crystalline core
at the central interstitial channel. 
The diamond structure wires with diameters smaller than 1 nm
suffered extensive reconstructions towards amorphous-like structures.
Among these, we mention a wire oriented
along (100), with a diameter of $\sim$ 0.9 nm, which after reconstruction
became corrugated with pentagonal rings at the surface. This wire is
shown in Fig.~\ref{nanowires2} (a).

{\it Fullerene-like nanowires - }
We considered two fullerene-based geometries proposed in
Ref.~\cite{fulerexp}, namely the Si$_{20}$ cage polymer and the
Si$_{24}$ cage polymer. We label those structures as ful3 and ful4,
respectively.  Based on the predicted stability of filled-fullerene-like
clusters ~\cite{parrinello}, we also considered variations of ful3 and
ful4, labeled as f-ful3 and f-ful4, with the inclusion of two extra atoms 
inside each cage. f-ful4 is shown in Fig.~\ref{nanowires2} (b).   
Its structure is corrugated and presents fivefold rings at the surface,
resembling the wire shown in Fig.~\ref{nanowires2} (a).
%hc Its corrugated structure and the fivefold rings at the surface
%hc makes it resemble the wire resulting from the reconstruction of the
%hc (100) cd wire, shown in Fig.~\ref{nanowires2} (a). 
For this reason, we
classify the latter as fullerene-like, and label it as f-ful5.
We further investigated filled fullerene-like nanowires of smaller diameter,
which we label as f-ful1 and f-ful2.
These are based on Si$_{12}$ and Si$_{16}$ cages, respectively,
with one additional atom in the  center of each cage.

{\it High-density nanowires - } The simple cubic (sc) Ge nanowire is shown
in Fig.~\ref{nanowires}(c). It shows very little distortions relative to
the bulk structure.
The $\beta$-tin nanowire has its axis parallel to the bulk
c-axis, passing through the center of an interstitial channel. While
the initial geometry of this wire is somewhat similar to that of the
sc nanowire, the corresponding relaxed geometries shown in
Fig.~\ref{nanowires} (c) and (d) are very different, due to the substantial
relaxation of the former. All four simple hexagonal (sh) nanowire 
structures are oriented 
along the bulk $c$ direction, and retain the crystalline order along 
the wire axis after geometry optimization, regardless of the wire radius. 
Two of the sh structures, sh1 and sh4, are shown in 
Fig.~\ref{nanowires}(e) and (f), respectively. 
We also considered an empty-hexagon 
variation of the sh1 structure in which the central atom of the hexagon
was removed. We label this structure as eh.
The relaxed Si
nanowires are structurally similar to the Ge ones shown in
Fig.~\ref{nanowires}.

\begin{figure}[h!]
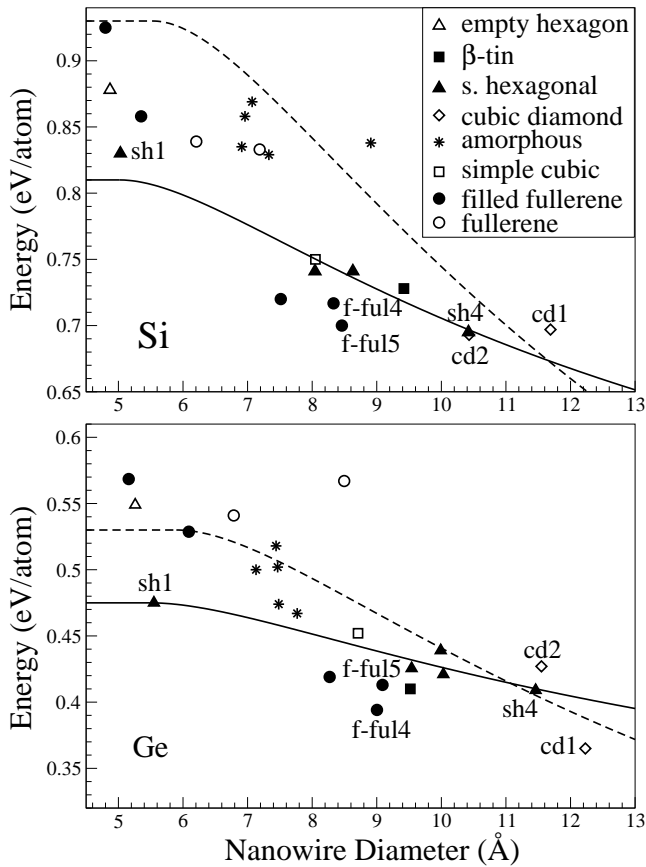

\centering
\includegraphics[clip]{si1.eps}
\includegraphics[clip]{ge1.eps}
\caption{Total energies per atom, $\Delta E_{tot}$ (in eV/atom,
relative to cd bulk energy) of Si and Ge nanowires
as a function of nanowire diameter. Labeled structures correspond to
those shown in Figs. 1 and 2. 
explained in the text. Lines show the curves obtained from the
continuum model (Eq. 1), with dashed (continuum) lines for nanowires based on 
the cd  (sh) bulk phase.}
\label{energy}
\end{figure}

%hc The results of our first-principles calculations for the total energy
The calculated total energies
of the nanowires are shown in the Fig.~\ref{energy}, where we plot
 $\Delta E_{tot}$ as
function of the nanowire diameter (defined as the diameter of the
smallest cylinder that contains the nanowire). We observe that the
formation energies of the high-density nanowires (based on sc, sh, and
$\beta$-tin phases) and the fullerene-like nanowires, with diameters 
of the order of 1 nm, are very
close to the energies of the diamond-structure nanowires cd1 and cd2, of similar
diameters, with energy differences of $\sim$0.05 eV/atom or less. Note
that these values are one order of magnitude smaller than the energy
differences of the corresponding bulk phases in Table~\ref{phases}.
This suggests that the energetics of wire formation in this diameter
range is strongly affected by the surface atoms~\cite{zhao03}.

Figure~\ref{energy} also shows that the amorphous wires derived from the
instabilities of thin, diamond-structure wires, have higher formation
energies than those of the high-density and the fullerene-like wires of
similar diameter. This suggests that amorphous wires could only be
produced in conditions very far from thermodynamical equilibrium.
Cage structures like the
unfilled fullerene structures and the empty-hexagon structures also have
very high energies as compared to denser structures.
The figure also shows that, among the nanowires of small diameters
(smaller than 0.7 nm), the sh1 structure appears below and to the left
in the energy vs. diameter diagram, which suggests a high stability
for this structure, when compared to the other small-diameter
geometries. In the diameter range between 0.7 and 0.9 nm, filled-fullerene-like
wires are the most stable ones. 

The above results suggest that the energetics of the formation
of nanowires are determined by the interplay between the energy per atom
of a ``bulk" part of the wire and the corresponding energy of a ``surface"
part of the wire. 
Motivated by that, we elaborated a simple continuum
model in terms of bulk and surface energies.  Let us consider a
cylindrical nanowire with radius R, number density $\rho$ (number of
atoms per volume), and total energy $E_{nw}$.  By decomposing $E_{nw}$
into a contribution due to the bulk-like atoms 
and a contribution due to the low coordinated
surface atoms, we write $E_{nw}=\varepsilon_{b} \rho V_{b}+\varepsilon_{s} \rho V_{s}$
%
%\begin{equation}
%E_{nw}=\varepsilon_{b} \rho V_{b}+\varepsilon_{s} \rho V_{s}
%\end{equation}
%
where $\varepsilon_{b}$ is the total energy per atom of the bulk-like
atoms, $\varepsilon_{s}$ is the total energy per atom of the surface
atoms, and $V_{b}$ ($V_{s}$) is the volume occupied by the bulk-like
(surface) atoms. $V_{b}$ and $V_{s}$ are determined by constraining the
surface to a monoatomic layer, such that the surface thickness becomes
$\rho^{-1/3}$. The total energy per atom, $\varepsilon_{nw}$, is then
written as
\begin{equation}
\varepsilon_{nw}=\frac{E_{nw}}{N_{at}}=\varepsilon_{s}+
(\varepsilon_{b}-\varepsilon_{s}) 
\frac{(R-\rho^{-1/3})^2}{R^2}
\label{nwenergy}
\end{equation}
for $R > \rho^{-1/3}$, where $N_{at}$ is the number of atoms of the nanowire.
For $R < \rho^{-1/3}$, $\varepsilon_{nw}=\varepsilon_{s}$.

To test our model, we apply Eq.~\ref{nwenergy} to nanowires based on
the sh and cd bulk phases of Si and Ge.  The parameters 
$\varepsilon_{b}$ and $\rho$ for the bulk
phases are obtained from our {\it ab-initio} results for each bulk phase.
The values of $\varepsilon_s$ for the sh and cd phases are obtained from 
best fits to the first-principles calculations of the nanowires.
In Table~\ref{model}, we show the fitted values of $\varepsilon_s$ for the
cd and sh phases of Si and Ge, as well as the difference 
$\Delta\varepsilon=\varepsilon_s-\varepsilon_b$ for each phase and
material. The Table shows that $\Delta\varepsilon$ is about
twice as large for the cd phase than for the sh phase for both Si and
Ge. This means that the energy cost of a cd surface is much larger than
that of a sh surface. This probably arises from the fact that surface
atoms in a cd structure are undercoordinated (coordination three or
less), while the surface atoms in a sh structure are still highly
coordinated, which reduces the energy cost of the surface.

\begin{table}
\centering
\caption{Surface energy relative to the cd bulk energy 
($\varepsilon_s-\varepsilon_b^{cd}$), in eV/atom, and 
relative surface energy within a given bulk phase
($\Delta\varepsilon=\varepsilon_s-\varepsilon_b$), in eV/atom,
for sh and cd-based nanowires  of Si and Ge.}
\vspace{0.30cm}
\label{model}
\begin{tabular}{lccc} \hline \hline
&                      
\multicolumn{1}{c}{$(\varepsilon_s^{sh}-\varepsilon_b^{cd})$}&
\multicolumn{1}{c}{$(\varepsilon_s^{cd}-\varepsilon_b^{cd})$}&
\multicolumn{1}{c}{$(\varepsilon_s^{sh}-\varepsilon_b^{sh})$}\\
\hline
%--------------------------------------------------------------------%
Si    &  0.810  &  0.930  & 0.415  \\
\hline                   
Ge    &  0.475 & 0.530 & 0.239     \\
\hline
\hline
%---------------------------------------------------------------------%
\end{tabular}
\end{table}

In Fig.~\ref{energy} we plot $\varepsilon_{nw}$ given by Eq. 1, relative to
the total energy per atom of the cd bulk phase, as a function of the
nanowire diameter for cd-based and sh-based nanowires of Si and Ge.
The first-principles results for the sh phase fall very near
$\varepsilon_{nw}$, which is a indication of the good quality of the
model. We also notice that, although $\varepsilon_s$ for the cd phase
was fitted only to the cd1 and cd2 wires (the ones that retained the cd
structure), the $\varepsilon_{nw}$ curve also
passes very near the calculated energies of the amorphous wires that 
originated from the cd structure. 

The results of the continuum model seen in Fig.~\ref{energy} 
clearly reproduce the trends of the
first principles results. In particular, the model correctly describes
the energy similarity of sh and cd nanowires for diameters around 1
nm. The model also predicts an inversion in the relative stability of
sh-based and cd-based nanowires for a diameter $D_c$ around 1.1 nm.
This stability inversion results essentially from the larger
$\varepsilon_s$ for the cd phase as compared to the sh phase, for both
Si and Ge. 
It is interesting to notice that the structural ``transition'' of the cd
nanowires, from crystalline-like (for d $ >$ 1 nm) to amorphous-like
(for d $<$ 1 nm), occurs at diameters very near the stability inversion
predicted by the model. This suggests that both effects are related, as
the cd phase becomes ``metastable'' for d $<$ 1 nm. 

Finally, let us comment on the stability of the filled-fullerene-like
structures and the sh1 structure for very small diameters. Although Eq.
1 cannot be applied to the filled-fullerene structures (there is not a
bulk structure associated with them), their calculated energies behave as
a decreasing function of the diameter. Such a
hypothetical curve would cross that of $\varepsilon_{nw}$ for the sh
phase for diameters near 0.7 nm, as one can see from Fig.~\ref{energy}.
This is consistent with the special stability of structure sh1 for very
small diameters, and it suggests that the sh phase might be the stable one
for ultra-thin nanowires.

In summary, we performed first-principles calculations of infinite,
periodic wires of Si and Ge with diameters between 0.45 and 1.25 nm.  
We found that the diamond structure nanowires are only stable
for diameters larger than 1 nm; thinner diamond-like
wires undergo considerable structural transformations towards
amorphous-like wires.
We propose a continuum model to explain the energetics of the
nanowires, on the basis of the competition between bulk and surface
energies. According to this model, parametrized from the
first-principles calculations, high-density nanowires become more
stable than the diamond structure wires for diameters smaller than 1 nm.
This is consistent with the structural instabilities
of the diamond-structure nanowires.
For diameters between 0.8 nm and 1 nm, filled-fullerene wires
are the most stable ones.  For even smaller diameters (d 
$\sim$ 0.5 nm), we find that the simple hexagonal structure is particularly
stable for both Si and Ge.

\begin{acknowledgments}

We acknowledge support from the Brazilian agencies CNPq, FAPEMIG,
and Instituto do Mil\^enio em Nanoci\^encias-MCT.

\end{acknowledgments}

\end{document}